\documentclass[12pt,epsf]{article}

\def\be{\begin{equation}}
\def\ee{\end{equation}}
\def\bea{\begin{eqnarray}}
\def\eea{\end{eqnarray}}

\def\be{\begin{equation}}
\def\ee{\end{equation}}
\def\bea{\begin{eqnarray}}
\def\eea{\end{eqnarray}}

\def\case#1/#2{\textstyle\frac{#1}{#2}}
\def\k0{\kappa_{0}}

\begin{document}
\begin{titlepage}

\vspace{.7in}

\begin{center}
\Large
{\bf Formation of a Black String in a Higher Dimensional Vacuum Gravitational Collapse}\\
\vspace{.7in}
\normalsize
\large{ 
 $Alexander Feinstein$
}\\
\normalsize
\vspace{.4in}

{\em Dpto. de F\'{\i}sica Te\'orica, Universidad del Pa\'{\i}s Vasco, \\
Apdo. 644, E-48080, Bilbao, Spain}\\
\vspace{.2in}
\end{center}
\vspace{.3in}
\baselineskip=24pt

\begin{abstract}
\noindent We present a solution to the  vacuum Einstein Equations which
represents a collapse of a  gravitational wave in 5 dimensions. Depending on the ``\emph{focal length}" of the wave the collapse  results, either in a black string covered by a horizon, or in  a naked singularity which can be removed. 
\end{abstract}

\vspace{.3in}

\end{titlepage}

There has been recently an  increasing interest  in higher dimensional solutions to Einstein field 
equations \cite{HD}. Such an interest stems mainly from the fact that the superstring description of the fundamental
interactions is necessarily described  by using a higher dimensional arena. Another reason to 
look at higher dimensional solutions
is because understanding these we might, as well, better
comprehend the four-dimensional world, its special character and uniqueness. On the other hand, we also start to appreciate  and learn the differences between the four and the higher dimensional physics.  

One of the most fascinating objects studied within the conventional General Relativity are the black holes. It happens so, that in higher dimensions, the study of the black ``objects" is not less relevant and it looks as the dynamics of the higher dimensional
black holes is richer in structure than their $4$-dimensional counterparts.

In four dimensions it is difficult, if not impossible, to construct an analytic solution to Einstein Equations describing  collapse of a gravitational wave into a black hole. To find such a solution, one presumably would need to know how to solve vacuum Einstein Equations with low symmetry. One can not employ the spherical symmetry  for it is well known that there are no spherical gravitational waves: these are prohibited by the Birkhoff theorem. Using the cylindrical symmetry in $4$-dimensions is also of no help. It is easy to convince oneself that cylindrical symmetry does not allow black holes. Lowering the symmetry of the problem results usually in impossibility to find the analytic solution, and one would  then be forced to rely on approximate or numerical methods for a highly non-linear situation. On a different venue, one could try  looking at collisions of strong plane waves, for example, but due to their infinite extent these produce rather naked singularities or unstable Cauchy Horizons \cite{planewaves}. The less symmetric, sufficiently strong p-p waves would certainly produce black holes upon collision \cite{d'Eath}, but analytically, again, there are no solutions describing such a collision.

In higher dimensions the situation is different. Apart from a round black hole, one may have  somewhat more ``cylindrical"  configuration, a black string. The black string solutions were studied in some details by now \cite{GregLafl}, but no analytic solution representing a  dynamical formation of such an object was presented so far. In this short note we present and initiate the study of this kind of solutions.

To this end we start with the following solution to $5$-dimensional  vacuum Einstein equations:

\be
ds^2=-(A/B)^{\sqrt{3}/3}\, dudv \,+ \, \frac{1}{4}\,(A/B)^{\sqrt{3}/3}\,A\,B\, d\Omega_{2} ^{2}\, + \,(A/B)^{-2\sqrt{3}/3}\, dw^2 \label{metric} 
\ee

Further generalization to higher dimensions is quite straightforward by changing the exponents of the $A/B$ factors of the line element and we do not discuss it here.

In the above equation
\be
A=\epsilon pv-bu, \qquad B=\epsilon  cv-du,
\ee
where $p,b,c,d$ are constants  satisfying $cb+dp=2$  and $\epsilon$ takes values $1$ or $-1$. The constant parameters$p,b,c,d$ introduce scaling of the coordinates $u$ and $v$ and should have dimensions of inverse length. In what follows we set $b=d=1$ and $p=1+a$, while $c=1-a$.  The $d\Omega_{2} ^{2}$ is a  two dimensional metric of a unit 2-sphere in the case $\epsilon =1$, while $\epsilon=-1$ corresponds to the case when the two-curvature is negative ($\sinh{\theta}$ replaces the $\sin{\theta}$). The $u$ and $v$ are  usual null coordinates. We will be working with the positive two-curvature $\epsilon=1$ mainly because we are interested in compact trapped surfaces. The line element \ref{metric}, if dimensionally reduced along the Killing direction $\partial / \partial{w}$ generalizes Roberts' solution \cite{Roberts} ( see as well \cite{Nakamura}).

 The radius of the two-dimensional sphere is given by
\be
r^2\,=\,\frac{1}{4}\,\left[A/B\right]^{\sqrt{3}/3}\,A \,B \label{radius}
\ee

Note, that the parameter $a$ defines the deviation of the line element from flatness and is related to  the strength of the wave. Indeed, the square of the Weyl tensor, which describes the focusing of the congruence of null geodesics \cite{penrose}, scales as \textbf{C}$^{2}  \,\propto a^2\,v^4$, therefore the parameter $a$ should be identified with  an \emph{inverse} of the focal length of the wave, analogous to those of the plane waves \cite{planewaves}. For $a=0$ we have Minkowski spacetime. The value $a=1$ separates the strong waves with short focal lengths ($a>1$) and the weak waves with long focal lengths ($a<1$).

Now, since the solution is vacuum, to find out  the curvature singularities   we look at the Kretschmann scalar $R^{acbd}R_{abcd}$ and come up  that it diverges as:
\be
 R^{acbd}R_{abcd}\propto \,A^{\frac{-12-2\sqrt{3}}{3}}  B^{\frac{-12+2\sqrt{3}}{3}},
\ee
therefore the curvature singularity is located at the origin $r=0$. This corresponds to $A=0$,
or in terms of $u$ and $v$, $u=(1+a)\,v,\,v<0$ and $B=0$ ($u=(1-a)v,\, v>0 $). The singularity is always timelike for $a<1$, but  becomes spacelike in the region $v>0$ along with $a>1$. We are not that interested in the case $a<1$, but rather in the case $a>1$ where the possibility of the spacelike singularity covered by an apparent horizon exists. The  simplest way to find the apparent horizons is to look at the  expressions worked out in \cite{senovilla} for trappedness of the surfaces. One may also look at the vanishing of the expansion scalar of the outgoing null geodesics. The scalar defining the trapping of some surface $S$, using the notations of the Ref. \cite{senovilla},  is given by :
\be
\kappa=-g^{ab}\,H_{a}H_{b},
\ee
evaluated on the surface. Therefore the apparent horizons (surfaces where the light wavefronts are instantly ``frozen")  are given by the solutions of the following equation:
\be
\kappa=-g^{uv}\,\frac{G_{u}}{G}\,\frac{G_{v}}{G}\,=0,
\ee
and $G=\frac{1}{4}A\, B\, \sin{\theta}$

Working out the expression for $\kappa$ we find that the locus of the points on the horizon is given by the equation :

\be
H :\Rightarrow \qquad u=(1-a^{2})\,v  
\ee

Therefore, the spacelike singularity at $r=0$ ($u=(1-a)v$, $v>0$) is shielded by the apparent horizon
at $u=(1-a^{2})\,v$, with the subsequent interpretation of the formation of the black string.

Because there exists an  unphysical timelike singularity in the region $v<0$, it is convenient to cut this region from the spacetime and paste instead  a flat metric. This can be done along the lines of the paper \cite{Nakamura}, see as well their Penrose's diagrams. While there is no scalar field to argue zero mass on $v=0$, it is possible to argue that the analoque of the energy flux as expressed in terms of Bel-Robinson superenergy or the Penrose's optical flux both vanish due to the vanishing derivatives of $\left(A/B\right),u$ on a $v=0$ surface. Note, however, that the line element fails to become asymptotically flat at future null infinity ($v\longrightarrow\infty$), and therefore again one must apply arguments of (\cite{Nakamura}) where they show that this sort of solutions are approximately valid in a strong field region $0 \leq v < v_{0}$.

Similar cutting and pasting exercise can be performed in the case $a<1$. Here, however we cut both the region $v<0$ and the region $u>0, v>0$ and paste there the flat spacetime to remove the timelike singularity. Therefore the interpretation of the spacetime is as follows:

We identify the parameter $a$ with the inverse focal length of the gravitational wave via the the $a$- dependence of the square of the Weyl tensor. The situation is similar to collision of plane gravitational waves, however, unlike in the plane wave collision, where the singularity, or a Cauchy horizon form independently of the strength of the waves due to their infinite extent, here, the strength of the wave plays a telling role. 

If $a<1$ the spacetime is asymptotically flat at the past null infinity $u\to-\infty$ and represents a collapse of a weak (long focal length) gravitational wave. The wave is weak enough so that after its passage the spacetime remains flat.

For $a>1$ (the short focal length ), the spacetime is still asymptotically flat at the past null infinity but the implosion produces a spacelike singularity  covered by the apparent horizon. Since the three-metric ($u=const\, , \, v=const$) is not round but rather looks as a string wrapped around the circle of a radius $\rho=(A/B)^{-\sqrt{3}/3}$ we interpret it as a black string. The string radius at the moment of the formation of the apparent horizon is: 
\be
\rho_{H}=\left[\frac{a-1}{a+1}\right]^{\sqrt{3}/3}
\ee

There are several aspects of the solution that are worth of exploring further. It is possible that the solution becomes unstable due to the so-called Gregory-Laflamme  instability (see the first reference in \cite{GregLafl}). In this case one would expect the break of the translational invariance in the extra dimension. The matching conditions on the null surface $v=0$ must be futher studied. It is possible that one must introduce null sheets of matter due to some discontinuities in the $\left(A/B\right),v$  derivatives on this surface. The higher dimensional generalisations of this solution do not change the behavior with respect to formation of the apparent horizon. It would be interesting, however,  to study the energetic and the entropy related issues in both $5$ and higher dimensions for this solution.  The study of Bekenstein-Hawking entropy in the context of the black string formation  may  shed further light on the notion of gravitational entropy. The quantum field theory in this spacetime is also of interest. Intuitively, the inverse of the focal length $a$  should relate to the temperature for the created particles, again, analogous to plane wave collisions \cite{verdaguer}.

\centerline{\bf Acknowledgments}
It is a pleasure to thank Roberto Emparan and Jose Senovilla for comments and correspondence.
This work is partially supported by the Basque government Grant GICO7/51-IT-221-07 and
The Spanish Science Ministry Grant FIS2007-61800.

\end{document}